# A Game Theory Based Ramp Merging Strategy for Connected and Automated Vehicles in the Mixed Traffic: A Unity-SUMO Integrated Platform

**Xishun Liao, Corresponding Author**
Center for Environmental Research and Technology
University of California, Riverside, Riverside, CA, USA, 92507
xliao016@ucr.edu

**Xuanpeng Zhao**
Center for Environmental Research and Technology
University of California, Riverside, Riverside, CA, USA, 92507
xzhao094@ucr.edu

**Guoyuan Wu, Ph.D.**
Center for Environmental Research and Technology
University of California, Riverside, Riverside, CA, USA, 92507
gywu@cert.ucr.edu

**Matthew J. Barth, Ph.D.**
Center for Environmental Research and Technology
University of California, Riverside, Riverside, CA, USA, 92507
barth@cert.ucr.edu

**Ziran Wang, Ph.D.**
InfoTech Labs
Toyota Motor North America, Mountain View, CA, USA, 94043
ziran.wang@toyota.com

**Kyungtae Han, Ph.D.**
InfoTech Labs
Toyota Motor North America, Mountain View, CA, USA, 94043
kyungtae.han@toyota.com

**Prashant Tiwari, Ph.D.**
InfoTech Labs
Toyota Motor North America, Mountain View, CA, USA, 94043
prashant.tiwari@toyota.com

**ABSTRACT**
Ramp merging is considered as one of the major causes of traffic congestion and accidents because of its chaotic nature. With the development of connected and automated vehicle (CAV) technology, cooperative ramp merging has become one of the popular solutions to this problem. In a mixed traffic situation, CAVs will not only interact with each other, but also handle complicated situations with human-driven vehicles involved. In this paper, a game theory-based ramp merging strategy has been developed for the optimal merging coordination of CAVs in the mixed traffic, which determines dynamic merging sequence and corresponding longitudinal/lateral control. This strategy improves the safety and efficiency of the merging process by ensuring a safe inter-vehicle distance among the involved vehicles and harmonizing the speed of CAVs in the traffic stream. To verify the proposed strategy, mixed traffic simulations under different penetration rates and different congestion levels have been carried out on an innovative Unity-SUMO integrated platform, which connects a game engine-based driving simulator with a traffic simulator. This platform allows the human driver to participate in the simulation, and also equip CAVs with more realistic sensing systems. In the traffic flow level simulation test, Unity takes over the sensing and control of all CAVs in the simulation, while SUMO handles the behavior of all legacy vehicles. The results show that the average speed of traffic flow can be increased up to 110%, and the fuel consumption can be reduced up to 77%, respectively.

## INTRODUCTION

Stated by Statista [1] in 2015, there were 1.28 billion motor vehicles in use in the world, and this number will grow to 2 billion within one or two decades. Traffic issues about safety, efficiency, and environment draw more attention as transportation is more involved in people's lives. Reported by the U.S. Department of Transportation, there were 36,560 people killed in traffic crashes on U.S. roadways in 2018 [2]. A survey was carried out by INRIX, showing that in 2019 traffic congestion costs each American nearly 100 hours, $1,400 a year, and the time lost in traffic jams had increased by 2 hours compared with 2017 [3]. Among the factors leading to traffic congestion and accidents, ramp merging has a significant impact [4)], due to the chaotic nature of driving behaviors and the lack of coordination in the merging area. As the emerging CAV technology enables vehicles to communicate with each other, many CAV-based algorithms and systems have been implemented to coordinate vehicle maneuvers or human driver behaviors in the merging area, such as centralized optimal control ramp merging [5][6], feedforward/feedback control ramp merging [7], and ramp merging speed guidance system [8], but most of them assume 100% penetration rate of CAVs.

However, since CAVs are supposed to share the road with legacy vehicles for a foreseeable future, the mixed traffic environment is more complicated, which makes it more challenging to regulate the entire traffic. The well-planned merging sequence or speed trajectories for CAVs may be interrupted by human-driven vehicles. Hence, the interaction between CAVs and legacy vehicles should be considered. In situations that involve human behavior, game theory has been recognized as one of the effective methods for decision making. Game theory can model how human decides to compete or cooperate, hence enables researchers to analyze the interaction between drivers [9]. This paper proposes a game theory-based ramp merging strategy for Connected and Automated Vehicles (CAV) in the mixed traffic, which is a decentralized agent-based algorithm and can provide the optimal merging sequence and respective speed trajectory for each CAV in real time.

Simulation is a widely used method to implement and evaluate the algorithms due to its characteristics of low-cost, low-risk and fast-feedback. Microscopic traffic simulators, such as VISSIM and SUMO, provide high fidelity and continuous traffic simulation to model complex vehicle interactions in a specified traffic network [10][11]. In game engine-based driving simulators, such as LGSVL and CARLA, researchers are provided with high degree of freedom and realistic simulation which enables the use of vehicle models, flexible specification of sensor suites, environmental conditions, and full control of all static and dynamic actors, maps generation [12][13]. In this research, we fuse the traffic simulator (i.e., SUMO), with the driving simulator (i.e., Unity), to verify the proposed strategy. In this Unity-SUMO integrated platform, CAVs are not only modeled with more realistic vehicle dynamics, but also equipped with customized radar systems and wireless communication modules. This integrated platform can also support the human-in-the-loop simulation. Various scenarios are used to evaluate the performance of the proposed strategy, such as different penetration rates and different traffic congestion levels, and many types of drivers are generated by SUMO to make the mixed traffic more realistic. Moreover, a real-world ramp merging area is mapped into both the Unity world terrain and SUMO traffic network.

Compared to other recent studies about highway ramp merging, the major contributions of this study are as follows:

- For a mixed traffic environment, a game theory-based ramp merging strategy for CAV is proposed, implemented and evaluated.

- A traffic flow level simulation is carried out on a self-built Unity-SUMO simulation platform, which provides high fidelity one-to-one mapped real-world terrain, vehicle model with sensing system, and traffic flow generator.

The rest of the paper is organized as follows. Section 2 discusses the background and related work. Section 3 illustrates the workflow of the system and the applied game theory algorithm. Section 4 presents the Unity-SUMO platform setup and evaluates the system performance at the traffic flow level. Conclusions and future work are presented in Section 5.

## BACKGROUND and RELATED WORK

This section discusses the background and recent work in ramp merging algorithms for CAVs, game theory-based ramp merging algorithms, the implemented acceleration control algorithm, and the current usage of the simulation platforms.

### Ramp Merging Algorithms for CAVs

A number of ramp merging strategies have been developed to increase the road safety and efficiency by leveraging CAV technology. Awal et al. proposed a proactive optimal merging strategy based on V2V communication to optimize on-ramp merging time and to reduce merging bottlenecks [13]. Baskar et al. proposed a centralized system to organize the connected vehicles into platoons via the communication between roadside controller and the merging vehicles, which conduct a dynamic speed limit for each vehicle using ramp metering [15]. Lu et al. introduced a concept of virtual platooning and proposed a close-loop adaptive longitudinal control algorithm to control the merging speed of CAVs [16]. Ahmed et al. proposed a decentralized algorithm for a freeway merge assistance system using Dedicated Short Range Communication (DSRC) technology, which provided a visual advisory on a Google map through a smart phone application [17]. Centralized algorithms are design to get a global perspective and obtain enough information, while distributed control algorithms are popular for the flexibility. In a mixed traffic, decentralized algorithms can be more robust in the mixed traffic, as it does not rely on the infrastructure and has a lower communication requirement than centralized system control.

### Game Theory-Based Ramp Merging Algorithms

Ramp merging is considered as mandatory lane changing with a geographical constraint, requiring both longitudinal and lateral control. To simplify the model, most of the research only focuses on the longitudinal control with assuming the vehicle can easily change the lane given enough inter-vehicle gap. Kita developed and calibrates a game theory model to describe the cooperative action that the mainline cars move to the adjacent lane to avoid the conflict with ramp car [18]. Liu et al. proposed a game describing how the vehicle on ramp and mainline choose to merge and yield [19]. A repeated game model was adopted by Kang and Rakha to make lane change decision with taking previous game results into account [20]. To get a global perspective and obtain an optimal solution, centralized optimization algorithms are developed to organize the ramp merging. Jing et al. proposed a cooperative game-based merging sequence coordinate system to arrange the CAVs into platoons, and used optimal control to guarantee the sequence [21]. Wang et al. combined receding horizon control with game theory to find an optimal acceleration control for both lane-changing and car-following [22]. However, a strong assumption of 100% CAV penetration rate rests in most of recent studies, allowing to build centralized complete game with full information. In contrast, especially in mixed traffic with low penetration rate, CAVs can only form an incomplete game with limited information from their surrounding legacy vehicles. Moreover, the advantage of CAVs' long range communication is diminished in mixed traffic, since long distance communication includes higher uncertainty of the environment.

### Consensus Based Acceleration Control Algorithm

In this study, we adopt the consensus control algorithm [23] from our previous research, which provides an acceleration control for ego vehicle to maintain a desire inter-vehicle gap and a same speed with its target vehicle. Ego vehicle needs to confirm its target and then obtain the real-time information of its target, including the longitudinal position and speed. Next, with this information, a reference acceleration $a_{ref}$ for ego vehicle is calculated by the consensus-based motion control algorithm, as depicted in the following equation:

$$a_{ref}(t+\Delta t) = -a_{ij}k_{ij} \cdot \left[\left(r_i(t) - r_j\left(t - \tau_{ij}(t)\right) + l_j + v_i(t) \cdot \left(t_{ij}^g(t) + \tau_{ij}(t)\right)\right) + \gamma_i \cdot \left(v_i(t) - v_j\left(t - \tau_{ij}(t)\right)\right)\right] \quad (1)$$

where $a_{ij}$ is the value of adjacency matrix for two vehicles; $k_{ij}$ and $\gamma_i$ are control gain parameters, respectively; $\tau_{ij}(t)$ is the time-variant communication delay between two vehicles; $t_{ij}^g(t)$ is the time-variant desired time gap between two vehicles; $l_j$ denotes the length of the leading vehicle; $v_j$ and $r_j$ are the longitudinal speed and the longitudinal position of the leading vehicle, respectively [23].

### Cross Platform Simulation

Oh et al. built a human in virtual reality in the loop simulator by connecting VISSIM and Unity with VR to create an immersive driving environment [24]. To study the vehicular communication, Amoozadeh et al. developed a Vehicular NeTwork Open Simulator (VENTOS), which combines SUMO and Objective Modular Network Testbed (OMNET++) and provides the functions of communication network and vehicular traffic simulator [25]. Biurrun-Quel et al. set up a driver-centric simulator by connecting Unity and SUMO through TraCI (Traffic Control Interface) protocol to present a driver view of one of the SUMO controlled NPC vehicles [26], which only allows one-way communication from SUMO to Unity for only presenting the driver's view. To simulate the mixed traffic and implement the algorithm for CAVs, the simulation platform is supposed to mimic the real world as much as possible, requiring but not limited to real world topography, vehicle dynamic models, vehicle perception systems, different driver behaviors and different traffic conditions.

## METHODOLOGY

This section elaborates the mixed traffic ramp merging strategy for CAVs. The system structure and strategy workflow are introduced in the first subsection, followed by the elaboration of the game theory algorithm formation and decision-making process.

### Strategy Workflow

Our strategy is designed from a decentralized agent-based model perspective, allowing vehicles to act independently. The strategy workflow is shown in **Figure 1**, and every vehicle goes through this process at each timestep. There are six major modules functioning to support the strategy.

1) The conflicts prediction module is designed to identify if potential conflicts exist, by consuming the information from the radar system. As **Figure 2** shows, it projects the ego vehicle and its detected surrounding vehicles into future to see if any surrounding vehicles will go in the safe distance of the ego vehicle. **Equation 2** describes how to identify the potential conflict during the interval of two timesteps, and then the conflicting vehicle will be added into a conflict list.

$$\begin{cases} \begin{matrix} v_i \times \Delta t + D_{safe} \leq v_j \times \Delta t + d_{ij} \\ or \\ v_i \times \Delta t - D_{safe} \geq v_j \times \Delta t + d_{ij} \end{matrix}, & No\ conflict \\ else, & Potential\ conflict \end{cases} \tag{2}$$

where $v_i$ is the speed of ego vehicle; $v_j$ is the speed of its surrounding vehicle;$d_{ij}$ is the intervehicle gap; $\Delta t$ is the simulation timestep length; $D_{safe}$ is the safe distance.

2) The conflict avoidance module is particularly designed for mainline vehicles since mainline vehicles can change to another lane to avoid the conflict with ramp vehicles. This module takes the time-to-collision (TTC) and inter-vehicle gap as the factors, urging mainline vehicles to change the lane for larger inter-vehicle gaps and TTCs. If mainline vehicles cannot change the lane, they activate the Game Formation module and play the game for the merging order with ramp vehicles.
3) Module 3, 4 and 5 consist of the game theory decision making process. The player identification module is designed to classify ego vehicle's potential competitors into either CAV or legacy vehicle, which determines the game type in the module 4. The module 3 contacts all radar detected vehicles by sending out an invitation, and all responsive and cooperative vehicles will be classified as CAV.
4) The game formation module is triggered once it receives the conflict list from the conflict prediction module, where ego vehicle forms a game with its competitors, each potentially conflicting vehicle. If the competitor is a CAV, the game is a cooperative game, otherwise, non-cooperative game. In the game, each player can choose to be a follower or a leader of its competitor. The corresponding acceleration and cost are calculated for each decision. Notably, the accelerations are computed by the consensus-based acceleration control algorithm from our previous study for vehicle's longitudinal control [23]. The inputs of the controller are the speed and position of the ego vehicle and its preceding vehicle, and the desired inter-vehicle gap, as shown in **Equation 1**.
5) The merging sequence determination module is the last part of game theory algorithm, besides the game formation module. This module is designed to coordinate the merging sequence dynamically by utilizing results from the game formation module. Each vehicle will obtain a role with regard to its competitor, leader or follower, as well as an optimal longitudinal acceleration that satisfies the safety constraints. If two competitors are both CAVs, they will share respective cost with each other and make decision together. The detail of game theory algorithm will be introduced in the next subsection.
6) The acceleration control module is designed for the driving simulator, and is responsible for two main goals. The first one is to ensure the vehicle can run at the desired longitudinal speed and track the lane. The second one is to perform the lane change maneuver safely, once the lane change condition is satisfied.

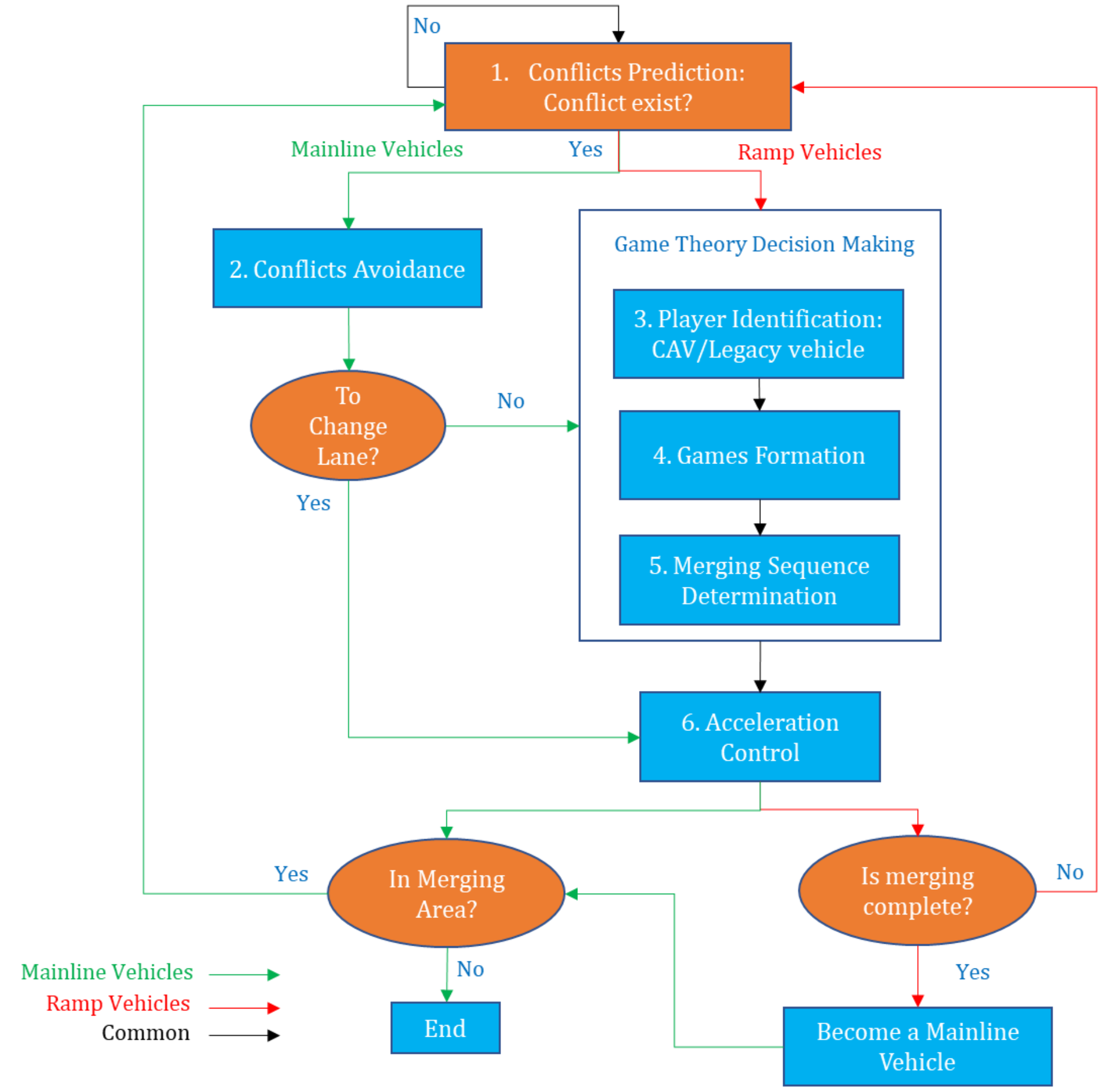

**Figure 1 System workflow of the mixed traffic ramp merging strategy for CAVs**

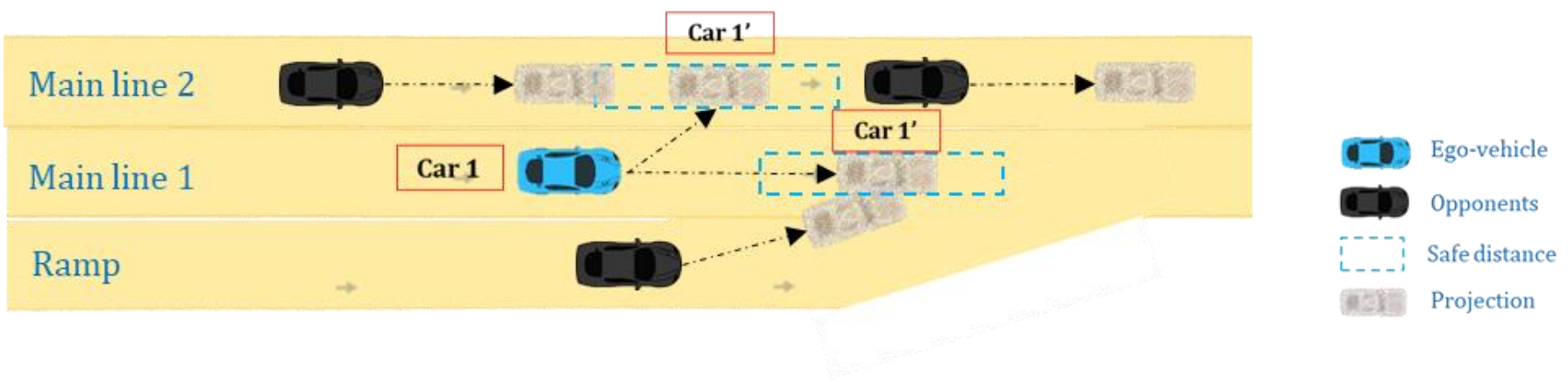

**Figure 2 Conflict prediction module**

## Game Theory-Based Merging Sequence Determination

*Game Formulation*

When the potential conflict exists in a merging area, at least one of mainline vehicle and ramp vehicle needs to adjust their speed for a certain merging sequence. These two competitors hence form a game unavoidably to evaluate their situation and decide who should go first. In this two-player game, the ego vehicle is named player 1 (P1), while its competitor is named player 2 (P2). Both P1 and P2 can choose either to be a leader or a follower, with the action set given as, A (P1) = {1: To be the leader, 2: To be the follower} for P1, and A (P2) = {1: To be the leader, 2: To be the follower} for P2.

Mainline vehicles have different motivation from ramp vehicles. Mainline vehicles attempt to drive safely without compromising in travel speed, while ramp vehicles have to worry about the remaining distance to the end of merging area. As the remaining distance decreases, the merging desire of the ramp vehicle may grow, and this anxiety can be expressed as the risk value in the cost function.

*Cost Function: Safety Term*

Safety is always the key factor to be considered. Kita [18] used the Time-to-collision (TTC) as the safety payoff to evaluate each decision. For each action in the game, a corresponding suggested acceleration $a$ is calculated by the control algorithm, hence we can predict the TTC in the next timestep of each action. The predicted TTC for any pair of players is formulated as **Equation 3**. However, using TTC only may not be enough to quantify the safety risk; for example, if the preceding vehicle is faster than the following vehicle, TTC will be negative. Moreover, if the difference between $v_f$ and $v_p$ is small, it will generate a huge TTC indicating safety even though the inter-vehicle gap is small. Kang and Rakha [20] proposed a safety assessment method to this problem by adding the inter-vehicle time headway into the cost function. By combining predicted TTC and predicted time headway, the risk for each action can be evaluated as **Equation 4**.

$$\mathrm{TTC}_{predict} = \frac{Gap + \Delta Gap(a, \Delta t)}{v_f + \Delta v_f - (v_p - \Delta v_p)} \qquad , if\, v_f + \Delta v_f > v_p - \Delta v_p \tag{3}$$

where $v_f$ is the speed of the following vehicle; $v_p$ is the speed of the preceding vehicle; and $\Delta t$ is the timestep.

$$h_{predict} = \frac{Gap+\Delta Gap(a,\Delta t)}{v_e+\Delta v_e}$$

$$risk_1 = \frac{\left(1 - \tanh\left(\frac{\mathrm{TTC}_{predict}}{t_h}\right) + 1 - \tanh\left(\frac{h_{predict}}{t_h}\right)\right)}{2}, \quad if v_e + \Delta v_e > v_c + \Delta v_c \tag{4}$$

$$risk_1 = 1 - \frac{\left(1 + tanh\left(\frac{h_{predict}}{t_h}\right)\right)}{2}, \quad if\ v_e + \Delta v_e < v_c + \Delta v_c$$

where $h_{predict}$,$v_e$, $v_c$ are the predicted time headway of ego vehicle, speed of ego vehicle, and speed of competitor vehicle, respectively; $t_h$ is the minimum safe time headway based on the 3-second rule [27]. There are two situations for $\mathrm{TTC}_{predict}$ computation: 1) if ego vehicle goes beyond its competitor, $\mathrm{TTC}_{predict}$ is the TTC of the competitor to ego vehicle; and 2) if ego vehicle goes behind its competitor, the $\mathrm{TTC}_{predict}$ is the TTC of ego vehicle to its competitor. The same rule applies to the calculation of $h_{predict}$.

To take into account the merging desire of ramp vehicles, the distance to the end of merging area should be added in the risk value of the ramp vehicle, as shown in **Equation 5**. The closer to the end, the higher the cost the vehicle should pay.

$$risk_{d2e} = \frac{1-\left(tanh\left(\frac{h_{ending}(t)}{t_h}\right)\right)}{2} \tag{5}$$

where $h_{ending}(t) = \frac{D_{end}}{v_{ramp}}$ is the remaining time headway to the end of merging area for the ramp vehicle; and $D_{end}$ is the remaining distance.

*Cost Function: Mobility Term*

As aforementioned, saving players' travel time is another target of the proposed strategy. If we only consider safety, being the follower in the game will always be the safest option. However, such conservative behaviors will result in unnecessary congestion, as the vehicles would prefer to slow down even though they may take the lead without exposing much safety risk. Adding a mobility term helps CAVs to find the balance between safety and speed, and improve the traffic efficiency at the same time. Both mainline and ramp CAVs are encouraged to take the action with less speed decrease. As shown in **Equation 6**, the mobility term adds more cost to a decelerating action. The term $tanh\left(\frac{\Delta v_e}{v_e}\right)$ is more sensitive when the speed of the ego vehicle is slow, as this algorithm cares more about mobility in low speed driving, but safety in high speed driving.

$$Mobility = \frac{\left(1 - tanh\left(\frac{\Delta v_e}{v_e}\right)\right)}{2} \tag{6}$$

where $\Delta v$ is the speed difference of either being a follower or leader in the game.
To summarize, the cost of mainline vehicle is:

$$Cost^{ML}(v_e, v_c, a, \Delta t) = risk_1^{ML} + Mobility^{ML} \tag{7}$$

The cost of ramp vehicle is:

$$Cost^{Ramp}(v_e, v_c, a, \Delta t, D_{ego,end}) = \frac{\left(risk_1^{Ramp} + risk_{d2e}\right)}{2} + Mobility^{Ramp} \tag{8}$$

*Non-Cooperative Game vs. Cooperative Game*

After obtaining the cost for each player's actions, the optimal result can be found in the decision table. The formulation of the decision table is based on the type of the game, which may be considered as either a non-cooperative or cooperative game. If the game is formed between a CAV and legacy vehicle, it would be a non-cooperative two-player game, where players make decisions on their own. The game between two CAVs would be a cooperative game, where players share their cost and make decisions together.

The decision table of the non-cooperative game between a CAV and legacy vehicle is shown in **Table 1.**

**Table 1: cost table for non-cooperative two-person game**

| | Competitor | | |
|---|---|---|---|
| Ego vehicle | Role | Leader | Follower |
| | Leader | $\infty$ | $Cost_{lead}^{Ramp}$ or $Cost_{lead}^{RAmp}$ |
| | Follower | $Cost_{follow}^{Ramp}$ or $Cost_{follow}^{ML}$ | $\infty$ |

To avoid collision, ego vehicle and its competitor are not allowed to be leader or follower at the same time, hence the cost for these two cases are set to be infinity or a very large number. At each timestep, ego vehicle will choose the option with the minimum expected cost, as described in **Equation 9.**

$$Action = \min_{actions}\left\{Cost_{Ramp}^{follow}\ or\ Cost_{ML}^{follow}, Cost_{Ramp}^{lead}\ or\ Cost_{ML}^{lead}\right\} \tag{9}$$

The decision table of the cooperative game between two CAVs is shown in **Table 2.** Unlike the non-cooperative algorithm which can only provide the local optimal solution due to the competition regardless of system conditions, a cooperative game for two CAVs can be solved with the cost table shown in **Table 2.** The cost of being a follower for ego vehicle $Cost_{follow}^{ego}$ is chosen from either $Cost_{follow}^{Ramp}$ or $Cost_{follow}^{ML}$ based on its position on road, and the same for the cost of competitor $Cost_{lead}^{com}$. Since two players can share the cost with each other, they are able to make decision together based on the sum cost of two players, as described in **Equation 10**, to achieve the system optimum.

**Table 2: cost table for cooperative two-person game**

| | Competitor | | |
|---|---|---|---|
| Ego vehicle | Role | Leader | Follower |
| | Leader | $\infty$ | $Cost_{lead}^{ego} + Cost_{follow}^{com}$ |
| | Follower | $Cost_{follow}^{ego} + Cost_{lead}^{com}$ | $\infty$ |

$$Action = \min_{actions}\left\{Cost_{follow}^{ego} + Cost_{lead}^{com}, Cost_{lead}^{ego} + Cost_{follow}^{com}\right\} \tag{10}$$

## CASE STUDY AND RESULTS EVALUATION

On the customized Unity-SUMO integrated platform, a traffic flow level simulation is carried out under different penetration rates and congestion levels. Then, the simulation results are analyzed in terms of mobility and fuel efficiency.

### Unity-SUMO Integration Simulation

As shown in **Figure 3(a)**, the upper part with terrain details is the Unity environment, and the lower part is the corresponding SUMO network. A two-way communication via UDP Socket connects and synchronizes these two simulation tools in real time, allowing SUMO to control the legacy vehicle while Unity controls CAVs with the proposed algorithm. A real-world network is coded in the simulation, spanning from the intersection of Chicago Avenue to the intersection of Iowa Avenue along Columbia Avenue in Riverside, California. It consists of one on-ramp and a segment of multi-lane mainline. A vehicle model of Lexus is used for both CAVs (in red) and legacy vehicles (in white), as shown in **Figure3 (b)**. The red rays spreading from the red CAV indicate the radar detection.

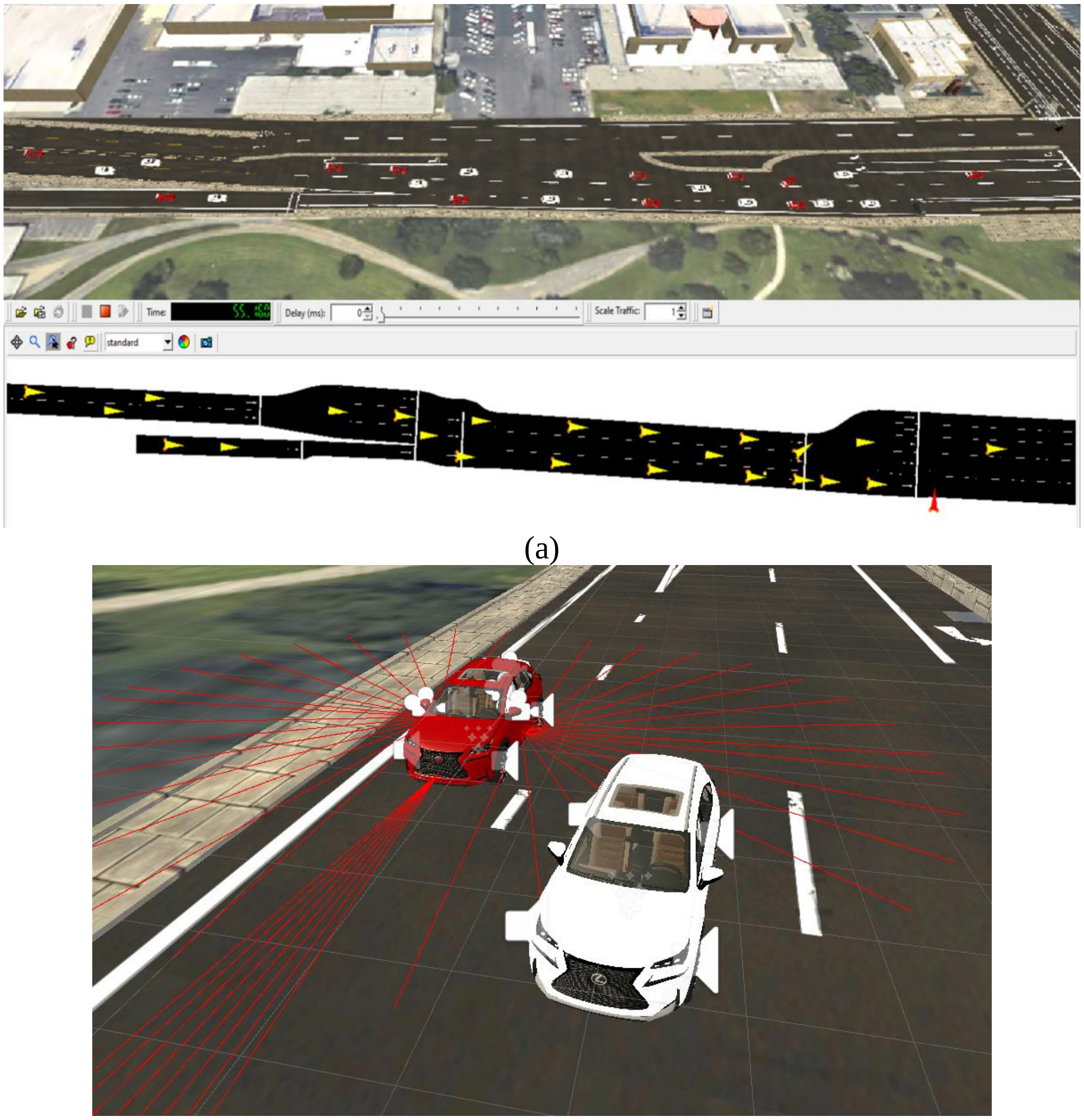

(a)

(b)

**Figure 3 Unity-SUMO integrated simulation.** (a) User interface of the simulation platform. (b) CAV (in red) with radar and legacy vehicle (in white).

## Simulation Design and Result Evaluation

### *Simulation Design*

To evaluate the performance of the proposed algorithm, the default control algorithms of SUMO are used as the baseline. In SUMO, the Krauss car-following model [28] is adopted for the longitudinal control, while the lane-changing model is developed by Jakob Erdmann [29]. The same control parameters are set in both the proposed algorithm and SUMO's algorithms, such as the desired time headway, desired speed, acceleration range, and minimum inter-vehicle gap, as shown in **Table 3**.

As shown in **Figure 3(a)**, vehicles are running on a two-lane mainline segment and a ramp, hence the mainline vehicles on the right lane can avoid the conflict with ramp vehicles by changing the lane. Based on the preset traffic demand, a traffic stream generator schedules the traffic itineraries, specifying each vehicle's departure time with Poisson distribution. As **Table 3** shows, the proposed algorithm is evaluated in three congested levels, including light traffic with the volume-to-capacity (V/C) ratio of 0.35, moderate traffic with the ratio of 0.6, and a congested traffic with the ratio of 0.85, whose flows are 1400, 2400 and 3400 vehicles per hour, respectively. In addition, to assess the performance of the proposed algorithm in various mixed traffic scenarios, four levels of CAV penetration rates (i.e., 0%, 30%, 70%, and 100%) are evaluated in the simulation. Therefore, in total 12 simulation runs with three congestion levels and four penetration rates are carried out.

**Table 3: Simulation Setup Parameters.**

| **Vehicular Parameters** | | |
|---|---|---|
| | CAVs | Legacy vehicles |
| Controller | Unity | SUMO |
| Initial speed (adaptive to traffic) | ramp: 15 m/s; mainline: 20 m/s | |
| Low speed minimum inter-vehicle gap | 5 m | |
| Acceleration range | $-5\sim3\ \mathrm{m/s^2}$ | |
| Desired speed | 20 m/s | |
| Desired time headway | 1 s | |
| Initial distance to merging point | ramp: 250 m; mainline: 280 m | |
| **Simulation Environment Parameters** | | |
| Penetration rate of CAVs | 0%, 30%,70%,100% | |
| Congestion level (vehicles/capacity) | 0.35, 0.60, 0.85 | |
| Demand (vehicles per hour) | 1400, 2400, 3400 | |
| Merging zone | 89 m | |
| Speed limit | 20 m/s | |
| Simulation timestep | 0.02 s | |
| Each simulation duration | 30 min | |

*Simulation Result Analysis*

For a fair comparison, we ensure all vehicles complete their trips and are cleared from the network at the end of each simulation run. The mobility and energy efficiency are analyzed for mainline and ramp vehicles, respectively, in each traffic scenario.

**Figure 4(a)** and **Figure 4(b)** present the average speeds (i.e., the ratio between vehicle-mile-traveled and vehicle-hour-traveled) of ramp vehicles and mainline vehicles, respectively. As shown in the figure, the proposed algorithm improves the average speeds for both ramp and mainline vehicles, especially for ramp vehicles. As shown in **Table 4**, the most significant average speed improvement is gained in the congested traffic, i.e., 109.59% for ramp vehicles and 31.98% for mainline vehicles, compared to the baseline. When the penetration rate grows, the average speed increases significantly because the proposed algorithm for each CAV can coordinate the merging maneuvers (including the sequence) implicitly, hence easing the congestion. Notably, in the scenarios with 100% penetration rate of CAVs, the average speeds for both ramp vehicles and mainline vehicles are close to the desired free flow speed (20 m/s), regardless of congestion levels. This means that the proposed algorithm can effectively regulate the traffic under different traffic demands when the penetration rate of CAVs is high.

**Table 4: Speed of traffic flow of each simulation.**

| **Ramp** | **Penetration rate** | **Baseline** | **30%** | **70%** | **100%** |
|---|---|---|---|---|---|
| Congested | Speed (m/s) | 9.07 | 13.61 | 17.47 | 19.01 |
| | Improvement (%) | 0 | 50.06 | 92.61 | 109.59 |
| Moderate | Speed (m/s) | 14.88 | 16.55 | 18.15 | 19.07 |
| | Improvement (%) | 0 | 11.22 | 21.98 | 28.16 |
| Light | Speed (m/s) | 17.84 | 18.52 | 19.07 | 19.11 |
| | Improvement (%) | 0 | 3.81 | 6.89 | 7.12 |
| **Mainline** | **Penetration rate** | **Baseline** | **30%** | **70%** | **100%** |
| Congested | Speed (m/s) | 14.32 | 17.28 | 18.65 | 18.90 |
| | Improvement (%) | 0 | 20.67 | 30.24 | 31.98 |
| Moderate | Speed (m/s) | 18.26 | 18.48 | 18.94 | 19.14 |
| | Improvement (%) | 0 | 1.20 | 3.72 | 4.82 |
| Light | Speed (m/s) | 18.94 | 18.98 | 19.07 | 19.24 |
| | Improvement (%) | 0 | 0.21 | 0.69 | 1.58 |

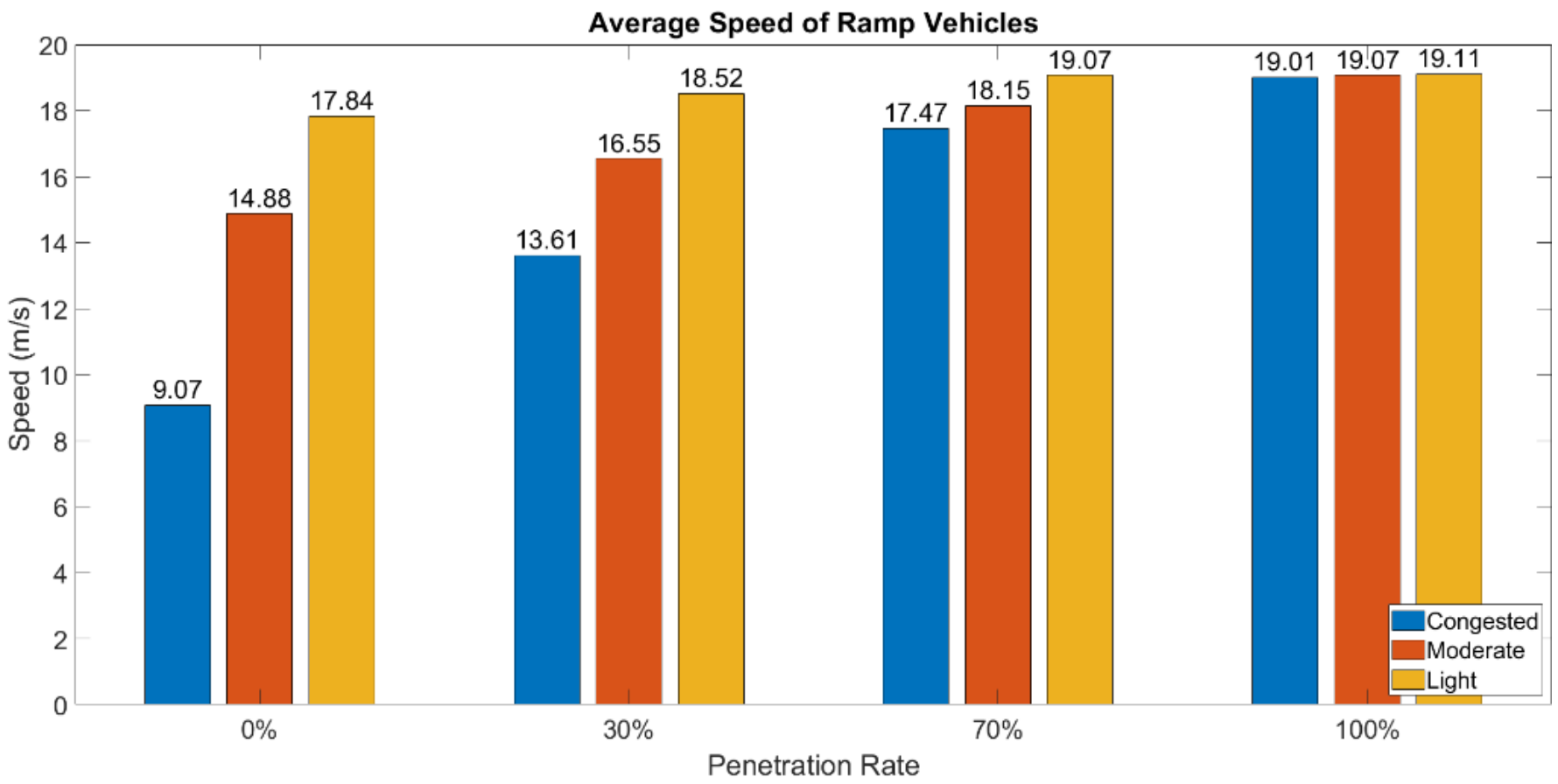

(a)

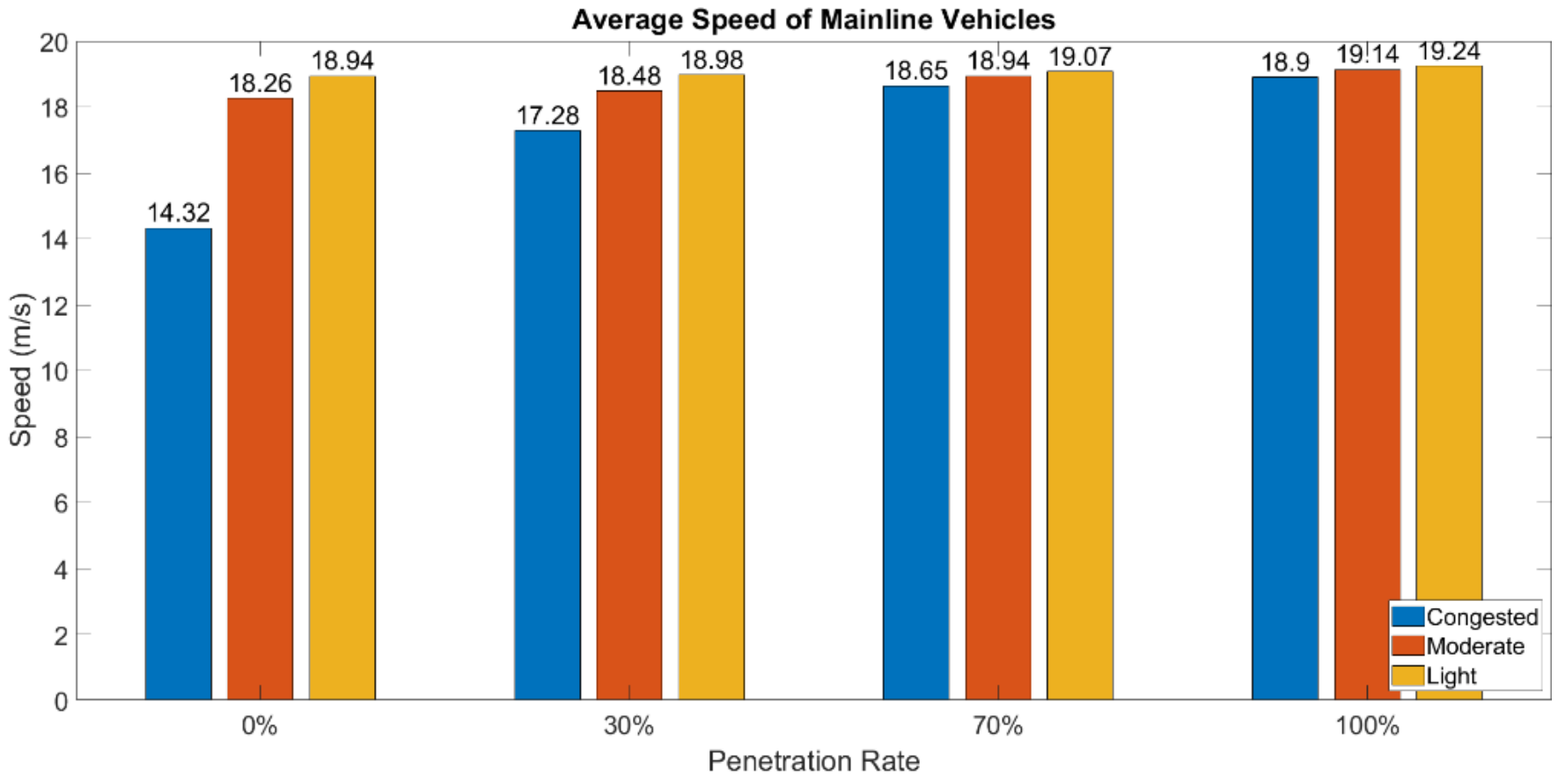

(b)

**Figure 4 Speed of traffic flow.** (a)Average speed of ramp vehicles. (b) Average speed of mainline vehicles.

We also analyze the fuel consumption with the U.S. Environmental Protection Agency's MOtor Vehicle Emission Simulator (MOVES) model [30]. **Figure 5(a)** and **Figure 5(b)** present the fuel consumption of ramp vehicles and mainline vehicles, respectively. As expected, results show that the proposed algorithm can considerably reduce the fuel consumption for both mainline and ramp vehicles, and the most significant reduction of fuel consumption happens in the congested traffic scenario for ramp vehicles. As shown in **Table 5**, up to 77.24% of fuel consumption can be saved since vehicles can merge in a coordinated manner and remove the stop-and-go behavior which is one of the major factors to generate the shock wave.

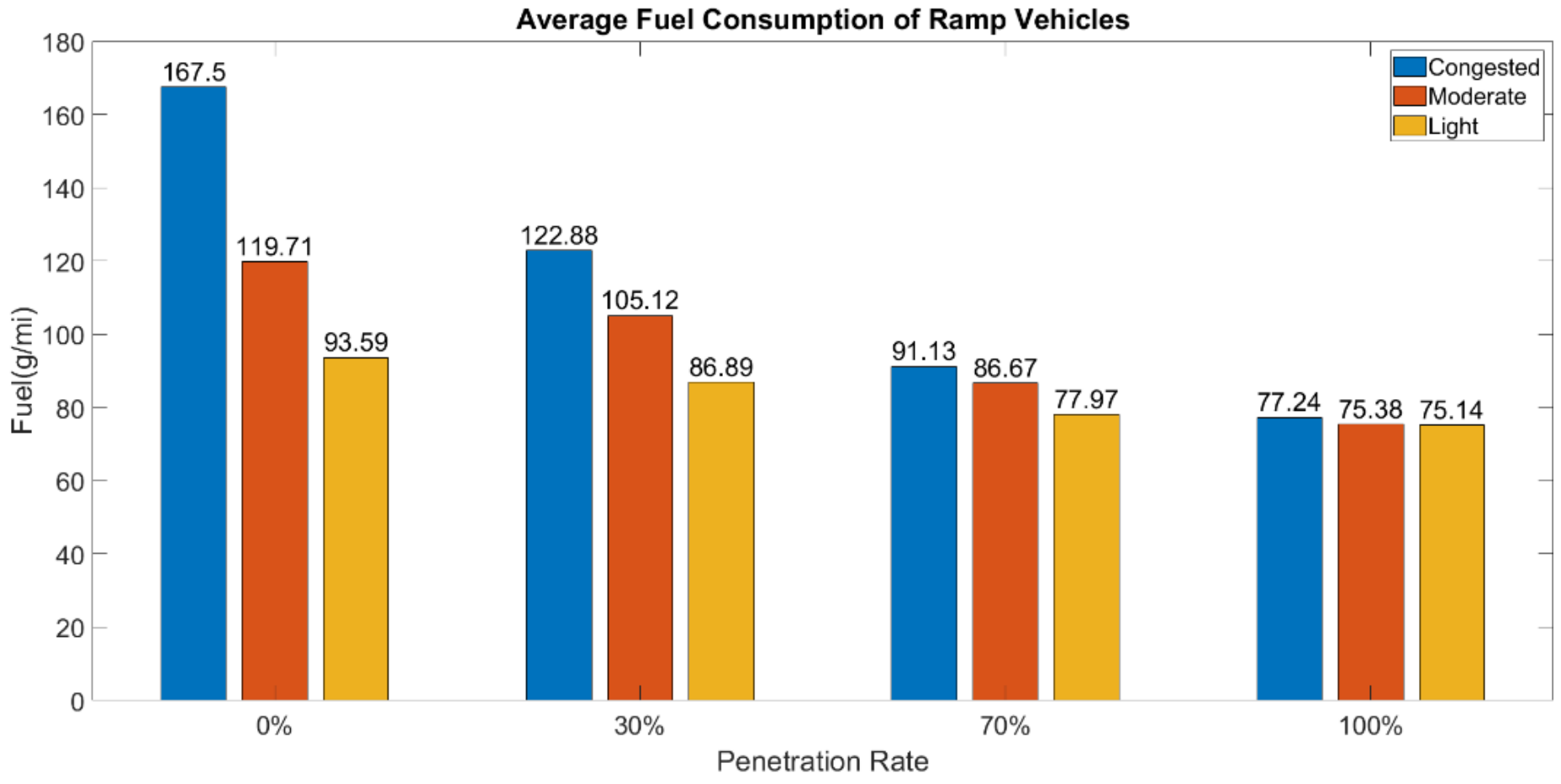

(a)

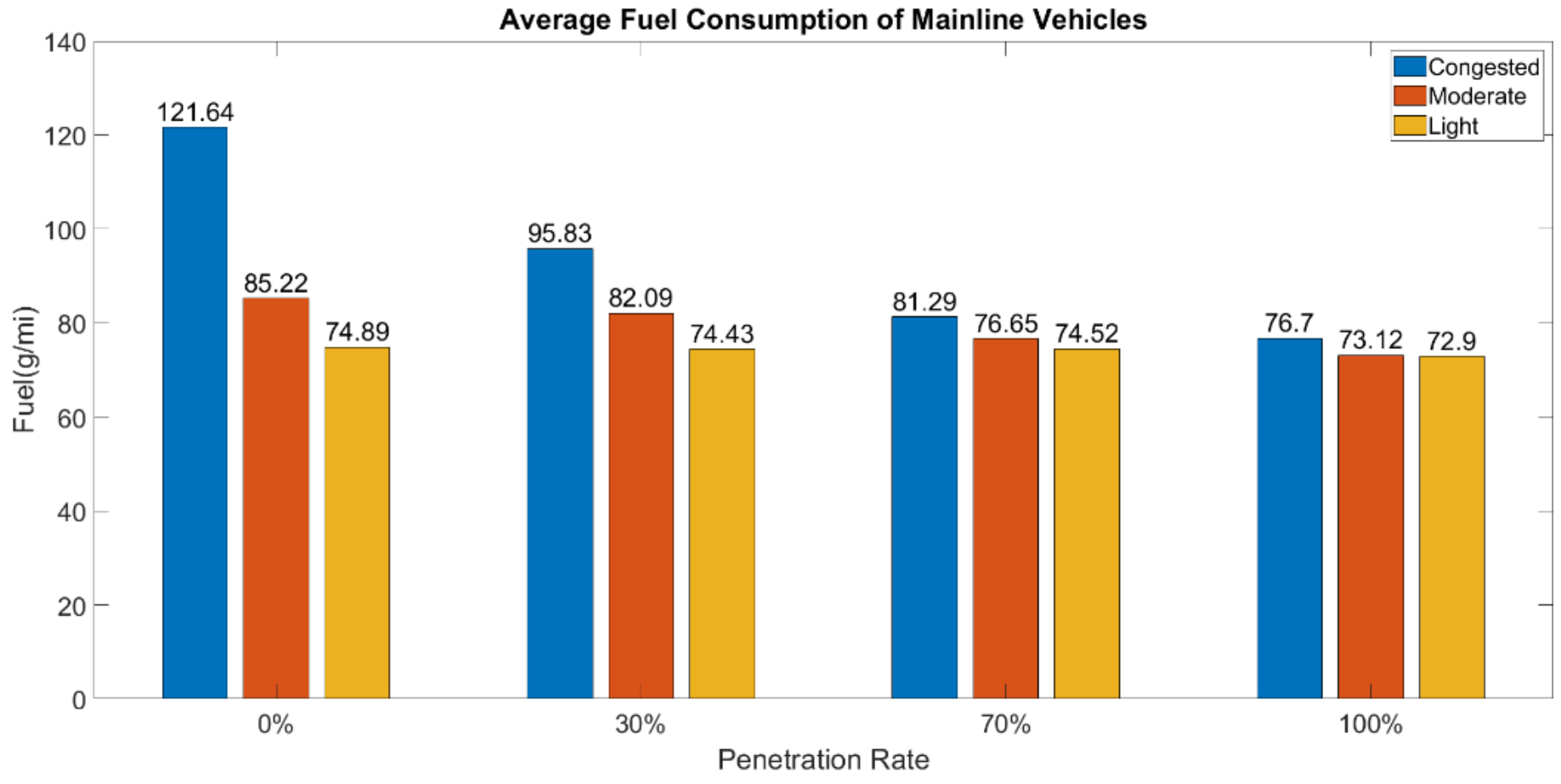

(b)

**Figure 4 Fuel consumption.** (a) Fuel consumption of mainline vehicles. (b) Fuel consumption of ramp vehicles.

**Table 5: Average fuel consumption of each simulation.**

| **Ramp** | **Penetration rate** | **Baseline** | **30%** | **70%** | **100%** |
|---|---|---|---|---|---|
| Congested | Fuel (g/mile) | 167.5 | 122.88 | 91.13 | 77.24 |
| | Reduction (%) | 0 | 26.64 | 45.59 | 53.89 |
| Moderate | Fuel (g/mile) | 119.71 | 105.12 | 86.67 | 75.38 |
| | Reduction (%) | 0 | 12.19 | 27.60 | 37.03 |
| Light | Fuel (g/mile) | 93.59 | 86.89 | 77.97 | 75.14 |
| | Reduction (%) | 0 | 7.16 | 16.69 | 19.71 |
| **Mainline** | **Penetration rate** | **Baseline** | **30%** | **70%** | **100%** |
| Congested | Fuel (g/mile) | 121.64 | 95.83 | 81.29 | 76.70 |
| | Reduction (%) | 0 | 21.22 | 33.17 | 36.95 |
| Moderate | Fuel (g/mile) | 85.22 | 82.20 | 76.65 | 73.12 |
| | Reduction (%) | 0 | 3.54 | 10.06 | 14.2 |
| Light | Fuel (g/mile) | 74.89 | 74.43 | 74.52 | 72.9 |
| | Reduction (%) | 0 | 0.61 | 0.49 | 2.66 |

## CONCLUSIONS AND FUTURE WORK

In this study, a game theory-based ramp merging strategy has been proposed for CAVs in a mixed traffic environment. The system has been developed, implemented and evaluated in a customized simulation platform, which fuses the functions of a game engine-based simulator, Unity and a traffic simulator, SUMO. Compared with the baseline merging algorithm of SUMO, the proposed algorithm can significantly improve the system mobility (up to 109.59%), and reduce the fuel consumption (up to 77.24%) under different traffic demands and CAV penetration rates

As one of the few ramp merging algorithms developed for the mixed traffic, numerous research problems need to be solved along its future development: 1) human-machine interaction in the game theory algorithm needs further investigation, since the modeling of human behaviors and the variation is still challenging; 2) with the development of the integrated Unity-SUMO simulation platform, human-in-the-loop simulation tests can be carried out for more in-depth human behavior study; and 3) besides the safety and mobility, the environment related term can be considered in the game function design to build an eco-friendly ramp merging system.

## ACKNOWLEDGMENTS

This research was funded by Toyota Motor North America, InfoTech Labs. We are grateful to Xiaofeng Zhang and Pingbo Ruan for providing the help before and during the Unity-SUMO platform building process.

The contents of this paper only reflect the views of the authors, who are responsible for the facts and the accuracy of the data presented herein. The contents do not necessarily reflect the official views of Toyota Motor North America, InfoTech Labs.

## AUTHOR CONTRIBUTIONS

The authors confirm contribution to the paper as follows: study conception and design: X. Liao, G. Wu, Z. Wang, and K. Han; construction of simulation platform: X. Liao, X. Zhao; data collection: X. Liao; analysis and interpretation of results: X. Liao, X. Zhao, G. Wu, Z. Wang, K.

Han; draft manuscript preparation: X. Liao, X. Zhao, G. Wu, M. J. Barth, Z. Wang, K. Han, P. Tiwari. All authors reviewed the results and approved the final version of the manuscript.